\documentclass[aps,pra,twocolumn,superscriptaddress]{revtex4}
\usepackage{graphicx}
\usepackage{amssymb}
\usepackage{subfigure}
\usepackage{amsmath}
\usepackage{amsfonts}
\usepackage{bm}
\usepackage{color}
\newcommand{\ket}[1]{\ensuremath{\left|{#1}\right\rangle}}
\newcommand{\bra}[1]{\ensuremath{\left\langle{#1}\right |}}

\newcommand{\oper}[1]{\mathbf{\mathsf{#1}}}

\begin{document}

%\preprint{}

\title{Experimental Observation of Quantum Correlations in Modular Variables}

%---------------------------------------
\author{M. A. D. Carvalho}
\affiliation{Departamento de F'sica, Universidade Federal de Minas Gerais, Caixa Postal 702, Belo Horizonte, MG
30161-970, Brazil}
%------------------------------------
%------------------------------------
\author{J. Ferraz}
\affiliation{Departamento de F'sica, Universidade Federal de Minas Gerais, Caixa Postal 702, Belo Horizonte, MG
30161-970, Brazil}
%------------------------------------
\author{G. F. Borges}
\affiliation{Departamento de F'sica, Universidade Federal de Minas Gerais, Caixa Postal 702, Belo Horizonte, MG
30161-970, Brazil}
%------------------------------------
\author{P.-L de Assis}
\affiliation{Departamento de F'sica, Universidade Federal de Minas Gerais, Caixa Postal 702, Belo Horizonte, MG
30161-970, Brazil}
\affiliation{Institut N\'eel, CNRS et Universit\'e Joseph Fourier, BP 166, F-38042 Grenoble Cedex 9, France}

%------------------------------------
\author{S. P\'adua}
\affiliation{Departamento de F'sica, Universidade Federal de Minas Gerais, Caixa Postal 702, Belo Horizonte, MG
30161-970, Brazil}
%---------------------------------
\author{S. P. Walborn}
\affiliation{Instituto de F\'{\i}sica, Universidade Federal do Rio de
Janeiro, Caixa Postal 68528, Rio de Janeiro, RJ 21941-972, Brazil}
\email{swalborn@if.ufrj.br}

\begin{abstract}
We experimentally detect entanglement in modular position and momentum variables of photon pairs which have passed through $D$-slit apertures.  We first employ an entanglement criteria recently proposed in [Phys. Rev. Lett. {\bf 106}, 210501 (2011)], using variances of the modular variables.  We then propose  an entanglement witness for modular variables based on the Shannon entropy, and test it experimentally.   Finally, we derive criteria for Einstein-Podolsky-Rosen-Steering correlations using variances and entropy functions.  In both cases, the entropic criteria are more successful at identifying quantum correlations in our data.  
\end{abstract}

%\keywords{}


\maketitle
\section{Introduction}
Quantum entanglement manifests itself in both discrete and continuous variable quantum systems.
Identifying and characterizing entanglement in continuous variable systems is quite challenging, due to the large Hilbert space structure and variety of measurements that can be performed.  This is especially true if the quantum state of interest is not Gaussian \cite{braunstein05,adesso07}.  Recently, an entanglement criteria was introduced for modular variables.  These correspond to the integer and remainder components of a continuous variable, such as the position $x$ or momentum $p$ \cite{gneiting11}.  Modular variables naturally arise in interference effects \cite{aharanov}, and provide a natural way to convert a continuous variable into a discrete component and a range-limited continuous component.   As such, they should find application in continuous-variable quantum information processing.  
\par
Entanglement witnesses using modular variables, such as that introduced in Ref.  \cite{gneiting11}, can identify entanglement in interesting quantum states. They are particularly well suited for bipartite states that describe non-local interference of two-particle wave packets.   A large series of experiments studying exactly this kind of non-local interference using entangled photon pairs has been performed over the last decade \cite{fonseca99b,fonseca00,neves05,peeters09}.  Using the spatial correlations produced by spontaneous parametric down-conversion (SPDC) \cite{walborn10}, it is possible to engineer a variety of two-photon entangled states by controlling the pump beam \cite{nogueira04,neves05,neves07,peeters09,walborn11a}.  In particular, if each photon passes through a $D$-slit aperture, the spatial correlations inherent in the SPDC process can be used to produce entangled $D$-dimensional qudits \cite{neves05}.  We also note that the spatial entanglement present in two-photon systems is also intricately related to quantum imaging and ghost diffraction \cite{walborn10,ribeiro94b,strekalov95,abouraddy01,santos08}.  
\par
Higher-dimensional quantum systems are interesting for quantum information tasks such as quantum cryptography \cite{bourennane01,cerf02} and bit commitment \cite{spekkens01}, and several studies have been performed for spatially entangled photons \cite{langford04,almeida05,walborn06a,walborn08b,zhang08}.  Qudits also present novel possibilities for fundamental tests of quantum theory \cite{collins02,amselem09,lapkiewicz11}, and to maximize the secure information capacity of entangled photons in the presence of experimental limitations \cite{leach12}.  
\par
Though there have been a number of experiments investigating aspects of entanglement in spatial qudits, useful entanglement witnesses have been lacking.  Previous experiments have used the conditional nature of the fringe pattern, which depends upon the spatial coordinates of both photons and only appears in the two-photon coincidence counts, to substantiate the non-separability of the system \cite{fonseca99b,fonseca99c,fonseca01}.   Discrete methods, based on the Schmidt decomposition, have been used to experimentally infer the entanglement in this system under the assumption that the state is pure \cite{neves07}.  
\par 
Here we produce pairs of photons in spatially entangled states of dimension $D \times D$ and use witnesses based on modular position and momentum variables to identify quantum correlations.   We first test an entanglement witness similar to that derived in Ref. \cite{gneiting11} to experimentally identify modular entanglement in the near-field ($x$) and far-field ($p$) variables, and obtain legitimate violation only for $D=2$.    Then, motivated by previous work in continuous variables \cite{walborn09}, we derive an entanglement  witness using the Shannon entropy function, and observe that it is more successful at identifying quantum correlations in our experimental results.  We further ask if the quantum correlations are sufficient to violate Einstein-Podolsky-Rosen-steering (EPR-steering) criteria for modular variables.   EPR-steering is a quantum correlation that is strictly stronger than entanglement \cite{wiseman07,saunders10}, and can be a resource for one-sided device independent quantum key distribution \cite{branciard12}.    
In section \ref{sec:steer}, we use the results of \cite{gneiting11} and our previous results to obtain EPR-steering inequalities based on variances and the Shannon entropy.  The entropic EPR-steering witness identifies entanglement in our data.    {An advantage of these witnesses is that they are valid for any initial bipartite quantum state.} 
 
  \section{Entangled Spatial Qudits}
Entangled qudits can be produced by sending spatially entangled photons through $D$-slit apertures \cite{neves05,neves07}.   In the near field of the slit apertures, the quantum state produced is well described as \cite{neves05}:
\begin{equation}
\ket{\Psi} = \frac{1}{\sqrt{D}}\sum\limits_{j=-l_D}^{l_D} \ket{\psi_j}_1\ket{\psi_{j}}_2,
\label{eq:slitstate}
\end{equation}
where $\ket{\psi_j}$ is the quantum state of a photon which passes through slit $j$ and $l_D=(D-1)/2$.  For $D=2$ these types of states have been used to identify a non-local de Broglie wavelength \cite{fonseca01} and investigate other fundamental aspects of non-local interference \cite{kim00,fonseca00,peeters09}.  These non-local interference patterns already indicate the presence of entanglement in the path variables of the prepared two-photon states. Our aim here is to develop a formal technique of detecting this entanglement.
\subsection{Experiment}
\label{sec:exp}
%%%%%%%%%%%%%%%%%%%%%%%
\begin{figure}
  \begin{center}
 \includegraphics[width=8cm]{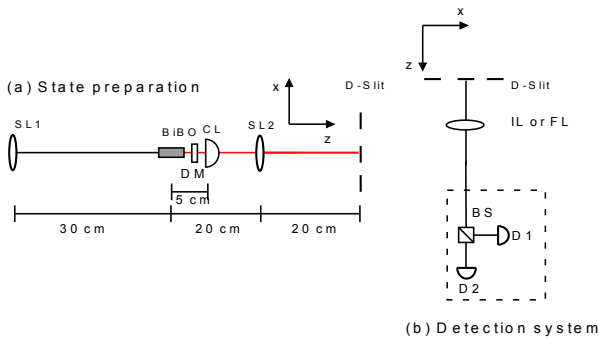}
  \caption{
(Color online). Experimental setup. (a) State preparation. A  CW laser generates collinear photon pairs in a BiBO crystal. Spherical lenses SL1, SL2 and cylindrical lens CL are used to increase the efficiency of  SPDC and image the spatial correlations of the photons at the D-slit (D = 2, 3, 4) plane, so that they both pass through the same slit of the D-slit aperture.  Dichroic mirror DM reflects the pump beam. (b) The detection system. The lens IL lens is used for near field imaging ($x$) and lens FL  for measurements in the far field ($p$).  BS is a $50/50 \%$ beam splitter and $D1$ and $D2$ are detectors.}
\label{fig:setup}
\end{center}
  \end{figure}
%%%%%%%%%%%%%%%%%%%%%%%
 We produced states of the form \eqref{eq:slitstate} experimentally.
 FIG. \ref{fig:setup} shows the experimental setup,  divided into two parts: state preparation in FIG. \ref{fig:setup}(a) and the detection system in FIG. \ref{fig:setup}(b). FIG. \ref{fig:setup}(a) illustrates the preparation of a spatially-entangled state similar to Eq. \eqref{eq:slitstate}.  A $ 50$ mW  continuous (CW) pump laser beam with $405$ nm wavelength crosses a $2$ mm BiB$_3$O$_6$ (BiBO) crystal and generates collinear photon pairs by type I SPDC. The photon pairs ($\lambda=810$nm) and pump beam propagate in the $z$-direction.  A dichroic mirror (DM) reflects the pump beam and transmits the down-converted photons.  The laser beam is focused in the center of the  crystal by a spherical lens (SL1) with $ 30$ cm focal length. After the crystal, a lens set is used to control the spatial correlation of the photons at the $D$-slit ($D= 2, 3,4$) plane \cite{peeters09,assis11}, located $40$ cm  from the center of the crystal. A magnified image of  the center of the crystal is projected at the $D$-slit plane by using a cylindrical lens (CL) with focal length $f_1 = 5$cm and a spherical lens (SL2) with focal length
$f_1 = 20$cm.  In this way, the correlation of the down-converted photons in the crystal is imaged on the $D$-slit aperture, which guarantees that the photons pass through the same slit, thus forming an entangled state \cite{peeters09,assis11}.  In the near field just after the $D$-slits, the photons can then be described by state \eqref{eq:slitstate}.   The width $a_D$ of each slit is $a_2=a_4=0.08$mm, $a_3=0.04$mm and their center-to-center separation $d_D$ is $d_2=d_4=0.16$mm and $d_3=0.125$mm.  The slits are placed perpendicularly to the $z$ direction,  with their smaller dimension in the $x$-direction. The length of the larger dimension of the slits ($y$-direction) is $8$mm, much larger than the down-converted beams, and can thus be considered to be infinite.  
\par
A detection system is set for imaging the $D$-slit plane (near field) or to project the Fourier transform of the $D$-slit (far field) at the detection plane, as shown in  FIG. \ref{fig:setup} (b). Both schemes use a spherical lens (IL or FL), a $50/50 \%$ beam splitter (BS) and two single-photon detectors  at the exit ports of the BS.  The detectors are placed $60$ cm from the $D$-slit plane, and equipped with interference filters centered at $810$ nm ($10$nm FWHM bandwidth) and $0.2$mm diameter pinholes. For the near field measurement, lens IL has a focal length $f_3 = 10$ cm and is placed 12.75cm from the $D$-slits, giving an image magnification of $M=3.6$, which was confirmed by independent measurement. For the far field detection, a spherical lens (FL) with a focal length $f_4 = 30$ cm in the $f$-$f$ configuration is used.    We toggle between measurement schemes only by switching lenses  IL and FL.  Each detector is mounted in a translation stage and can be scanned in the $x$ direction. Coincidences between the detectors are obtained with a homemade coincidence detection system with $5.4$ns time window. 
%%%%%%%%%%%%%%%%%%%%%%%
\begin{figure}
  \begin{center}
\includegraphics[width=8cm]{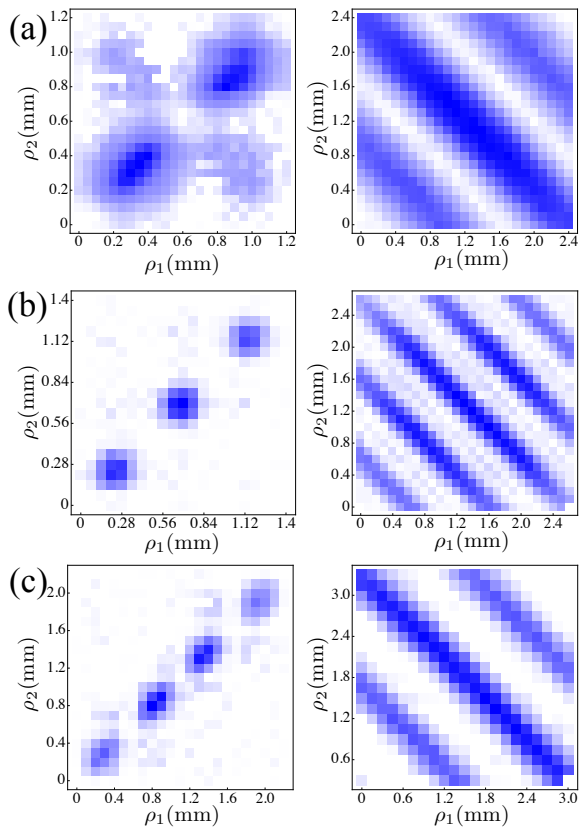}
  \caption{
(Color online.) Coincidence maps for (a) $D=2$, (b) $D=3$ and (c) $D=4$ slits.  $\rho_j$ ($j=1,2$) is the position of detector $Dj$. Graphics on the left correspond to near-field ($x$) measurements, showing correlation at the slit planes:  the photons pass through the same slit.  Graphics on the right are the far-field ($p$) measurements, showing fringes from non-local interference between the two-photon wavepackets.}
\label{fig:results}
\end{center}
  \end{figure}
%%%%%%%%%%%%%%%%%%%%%%%
\par
Two-dimensional arrays of coincidence counts were obtained by scanning both detectors in the $x$-direction of either the near-field plane or the far-field plane.  FIG. \ref{fig:results} shows the coincidence distributions for $D=2,3,4$ slits obtained in $2$ seconds (near field) or $30$ seconds (far field) for each point. Each coincidence array has more than $20 \times 20$ detection points.  For each near and far-field measurement, the detection position $\rho_x$ and $\rho_p$ in the $x$-direction of each detector was recorded, respectively.   The position variables are then given by $x=\rho_x/M$, and the momentum variables by $p=\rho_p/f\lambda$.  The corresponding two-dimensional probability distributions were obtained by normalizing the coincidence distributions.   These probability distributions will be used to evaluate the entanglement and EPR-Steering criteria using modular position and momentum variables presented in the following sections.      
\section{Modular Variables}
\par
Let us first define the modular variables for our two-photon system.  We will then present the entanglement criteria proposed in Ref. \cite{gneiting11}. We consider position and momentum operators that obey the commutation relation $[\oper{x},\oper{p}]=i$. Choosing a scale factor $\ell$ with dimension of length, one can define modular variables for the continuous $x$ and $p$ variables \cite{gneiting11}:
\begin{subequations}
\label{eq:modvar}
\begin{align}
x & = n \ell + r, \\
p & = m \frac{1}{\ell} + s,
\end{align}
\end{subequations}
where $n$ is the integer component of $x/\ell$ and $r=(x+\ell/2) \mod \ell - \ell/2$ is defined so that $-\ell/2 < r < \ell/2$.  Similarly, $m$ is the integer component of $p \ell$ and $s=(p+1/2\ell) \mod 1/\ell - 1/2\ell$.
For the two-photon system we can define the global modular variables 
\begin{subequations}
\label{eq:modvarD}
\begin{align}
{N}_{\pm} & = {n}_{1} \pm {n}_{2}, \\
{M}_{\pm} & = {m}_{1} \pm {m}_{2},
\end{align}
\end{subequations}
for the integer components and
\begin{subequations}
\label{eq:modvarC}
\begin{align}
{R}_{\pm} & = {r}_{1} \pm {r}_{2}, \\
{S}_{\pm} & = {s}_{1} \pm {s}_{2},
\end{align}
\end{subequations}
for the modular remainders.
\par
The variances of the distributions describing the modular variables ($j=1,2$) satisfy the uncertainty relation \cite{gneiting11}
\begin{equation}
\langle\Delta^2 {n}_{j} \rangle + \ell^2 \langle\Delta^2 {s}_{j}  \rangle \geq C=0.078235\dots, 
\label{eq:uncrel}
\end{equation}
where the constant $C$ is obtained numerically by calculating the smallest eigenvalue of the operator $\oper{n}_{j}^2 + \ell^2 \oper{s}^2_{j}$.
Using inequality \eqref{eq:uncrel}, one can show that for any bipartite separable state $\varrho=\sum_{i}p_i \varrho_{1i} \otimes \varrho_{2i}$, the following inequality holds \cite{gneiting11}:
\begin{equation}
\langle\Delta^2 {N}_{\pm} \rangle_\varrho + \ell^2 \langle\Delta^2 {S}_{\mp} \rangle_\varrho \geq 2C.
\label{eq:entcrit}
\end{equation}
Thus, violation of this inequality indicates that the state is entangled.  For state \eqref{eq:slitstate}, one can calculate
$\langle\Delta^2 {N}_{-} \rangle_\Psi=0$ and  $\ell^2 \langle\Delta^2 {S}_{+} \rangle_\Psi = \ell^2/6-(\ell^2/\pi^2)\sum_{j=1}^{D-1}(D-j)/Dj^2$, indicating a theoretical violation of the entanglement witness \eqref{eq:entcrit} for $D\geq 2$ \cite{gneiting11}.
\subsection{Experimental test of Variance Criteria}
%%%%%%%%%%%%%%%%%%%%%%%
\begin{figure}
  \begin{center}
\includegraphics[width=8cm]{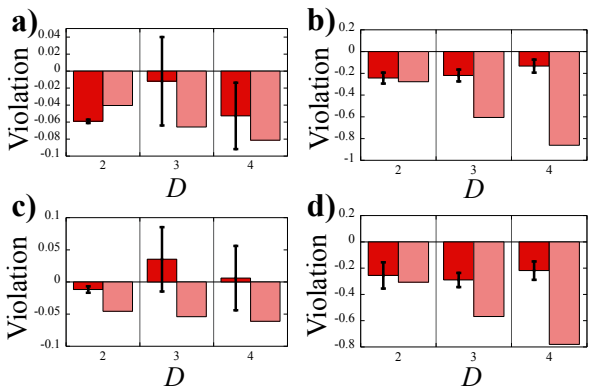}
  \caption{
(Color online.) Violation of entanglement and EPR-steering criteria for  $D=2, 3, 4$ slits.  Violation is defined as the right-hand side of the inequality subtracted from the left-hand side, so that negative values correspond to violation of the criteria and identification of quantum correlations.  The dark red bars show the experimental values, and the light red bars are the theoretical predictions calculated from the ideal state \eqref{eq:slitstate}. Figure a) shows the variance entanglement criteria \eqref{eq:entcrit}, figure b) the entropic entanglement witness \eqref{eq:uncrelNs+} , c) the variance EPR-steering criteria \eqref{eq:steercrit} and d) the entropic EPR-steering criteria \eqref{eq:entsteer}. For both entanglement identification and EPR-steering, the entropic criteria are more successful with our data.}
\label{fig:results2}
\end{center}
  \end{figure}
%%%%%%%%%%%%%%%%%%%%%%%

Using the experimental data presented in section \ref{sec:exp} and shown in Figure \ref{fig:setup}, we tested the entanglement criteria \eqref{eq:entcrit}. The scale factor $\ell$ is identified as the slit separation:  $\ell = d_D$.   
The variances in $N_-$ and $S_+$ used in the entanglement witness \eqref{eq:entcrit} were calculated using these probability distributions and the definitions of the modular variables \eqref{eq:modvar}, \eqref{eq:modvarD} and \eqref{eq:modvarC}.  The violation of the criteria (defined as the left-hand side subtracted from the right hand side of inequality \eqref{eq:entcrit}) is represented by the red bars in  
FIG. \ref{fig:results2} a).  In all results presented the error bars correspond to one standard deviation (SD), obtained by error propagation of the Poissonian count statistics.  Significant violation ($> 3$ SD) of entanglement criteria \eqref{eq:entcrit} is obtained only for $D=2$ slits.   The shaded red bars show the theoretical prediction for the entangled state \eqref{eq:slitstate}.  For $D=3,4$, the count rates were lower than for $D=2$ slits, as can be noticed from the error bars, and the noise (less than 5\%) in the near-field measurements prohibited reliable identification of entanglement with the variance criteria  \eqref{eq:entcrit}, and is responsible for the discrepancy between the experimental values and theory.  The far-field variance $\langle\Delta^2 {S}_{+} \rangle_\Psi$ was close to the theoretical prediction in all cases.      
The mediocre success of the variance criteria for our data motivates the search for  a more sensitive entanglement witnesses.  In the next section, we present such a criteria.  
%%%%%%%%%%%%%
 \section{Entropic Entanglement Witness for Modular Variables}
 It is known that the variance is not the optimal estimate of uncertainty \cite{cover}.  This has led to the development of entropic uncertainty relations \cite{bialynicki75}, as well as witnesses for entanglement \cite{giovannetti04,walborn09,saboia11}  that are based on entropy functions.  These witnesses typically outperform variance-based witnesses in the case of continuous variable non-Gaussian states.  Here we will derive an entropic entanglement witness for modular variables.  Our derivation will follow that of Refs. \cite{walborn09,saboia11}.   We will first consider pure separable states, and then extend our results to mixed states.  Consider a separable pure state
 \begin{equation}
 \Phi_{sep}(p_1,p_2) =\Phi_1(p_1) \Phi_2(p_2), 
 \label{eq:sepstate}
 \end{equation}   
  from which one can obtain the discrete probability distributions $P(n_1)$ and $P(n_2)$, as well as the probability densities $p(s_1)$ and $p(s_2)$ for the modular variables defined in Eqs. \eqref{eq:modvar}. In appendix \ref{sec:uncposmom} we show that the modular variables obey the entropic uncertainty relation
 \begin{equation}
 H(n_j) + h(s_j) \geq \ln \frac{1}{\ell},
 \label{eq:entuncrel}
 \end{equation}
 where $H(n)$ and $h(s)$ are the discrete and continuous Shannon entropies \cite{cover}.   From state \eqref{eq:sepstate}, we can calculate the probability densities $p(S_{\pm})$ for global variables $S_{\pm}$ defined in Eq. \eqref{eq:modvarC} as 
 \begin{equation}
 p_{\pm}(S_{\pm}) = p_{1} * p_{2}^{(\pm)},
 \end{equation}
 where $p_j=p_j(s_j)$, $p_j^{(\pm)}=p_j(\pm s_j)$, and ``*" denotes the convolution operation.   A continuous variable whose probability density is the convolution of two probability densities satisfies the entropy power inequality \cite{cover}, and we can write
 \begin{equation}
h(S_{\pm}) \geq \frac{1}{2} \ln \left \{ \exp[2h(s_1)]+\exp[2h(s_2)] \right \}. 
\label{eq:entsumS}
\end{equation}
To derive an entanglement witness following Refs. \cite{walborn09,saboia11}, we need a similar inequality for the discrete variables $N_{\pm}$.  Let us assume that variables $n_j$ are non-zero for some finite set of values, with corresponding discrete probability distributions $P_j$.  Since the quantum state \eqref{eq:sepstate} is pure and separable, variables $n_1$ and $n_2$ are independent.  Then, one can show that $H(N_{\pm}) \geq H(n_j)$ \cite{cover}.  That is, the sum or difference of two independent variables increases uncertainty.  It then follows that 
\begin{equation}
H(N_{\pm}) \geq \frac{1}{2} \ln \left \{ \frac{1}{2} \exp[2 H(n_1)] + \frac{1}{2}\exp[2 H(n_2)]\right \}.
\label{eq:entsumN}
\end{equation}    
We note again that inequality \eqref{eq:entsumN} is valid only for independent variables $n_1,n_2$.  Adding inequalities   \eqref{eq:entsumS} and  \eqref{eq:entsumN}  gives
 \begin{equation}
H(N_{\pm}) + h(S_{\pm}) \geq \frac{1}{2} \ln \left(\frac{1}{2}\sum\limits_{ij=1,2} e^{2 H(n_i) + 2 h(s_j)} \right).
\label{eq:entsumNS}
\end{equation}    
Using the uncertainty relation \eqref{eq:entuncrel}, it is straightforward to show that
 \begin{equation}
H(N_{\pm}) + h(S_{\pm}) \geq \ln \frac{\sqrt{2}}{\ell}.
\label{eq:uncrelNs+}
\end{equation}   
We note that \eqref{eq:uncrelNs+} actually describes four inequalities, all of which are satisfied by pure separable states.     
 \par
 To extend these inequalities to mixed states, we note that any bipartite separable state can be written as a convex sum of bipartite separable pure states
 \begin{equation}
 \varrho_{sep} = \sum_w \lambda_w \ket{\Phi_{w}}\bra{\Phi_{w}}, 
 \end{equation}
 where $\lambda_w \geq 0$ and $\sum_w \lambda_w=1$. 
  From the concavity of the Shannon entropy and inequality \eqref{eq:uncrelNs+} it then follows that
    \begin{align}
H(N_{\pm})_{\varrho} + h(S_{\pm})_{\varrho} & \geq \sum_w \lambda_w \left \{ H(N_{\pm})_{\Phi_w} + h(S_{\pm})_{\Phi_w}\right \}, \nonumber  \\
& \geq \ln \frac{\sqrt{2}}{\ell}.
\label{eq:uncrelNs+w}
\end{align}   
Thus, inequalities \eqref{eq:uncrelNs+}  are satisfied by any bipartite separable state.  However, two of these are satisfied by all bipartite states. This follows from the fact that operators corresponding to $N_+$ ($N_-$) and $S_+$ ($S_-$) do not commute.  However, it is possible to find entangled states for which $H(N_{+}) + h(S_{-}) < \ln ({\sqrt{2}}/{\ell})$ or $H(N_{-}) + h(S_{+}) < \ln ({\sqrt{2}}/{\ell})$.   We thus have an entropic entanglement witness for modular variables. As an example, consider state \eqref{eq:slitstate}, for which one can calculate $H(N_-)=0$ for all $D$.  For $D=2,3,4$, we calculate $h(S_+) - \ln ({\sqrt{2}}/{\ell})= -0.28, -0.61, -0.86$, respectively, showing increasing violation of criteria \eqref{eq:uncrelNs+} with $D$.  We note that one is free to choose the scale factor $\ell$.  In appendix \ref{sec:lmod}, we perform a numerical investigation of $\ell$, and show that $\ell=d$ (the slit separation) is the optimal choice.  
 \subsection{Experimental Test of Entropic Entanglement Witness}
 \par
Using the probability distributions associated to coincidence counts in FIG. \ref{fig:results}, we calculated the entropic modular entanglement witness \eqref{eq:uncrelNs+} for $N_-$ and $S_+$ variables.  Violation of the criteria is shown by the solid red bars in FIG. \ref{fig:results2} b).  We obtain a significant ($>3$SD) violation for all values of $D$, confirming that entanglement is present in all cases.   The shaded red bars show the theoretical prediction for the entangled state \eqref{eq:slitstate}.  We note a considerable discrepancy between experiment and theory, particularly for $D=3,4$.  This is due primarily to the noise in the near-field measurements, and also to the fact that the probability distribution for $N_-$ is not constant, as it is for the ideal state \eqref{eq:slitstate}.  Nonetheless, the entropic entanglement witness  \eqref{eq:uncrelNs+} successfully identifies entanglement.  
      
\section{EPR-Steering Criteria for Modular Variables}
\label{sec:steer}
One can also ask whether the correlations present in the state \eqref{eq:slitstate} and/or the experimental data present any stronger form of quantum correlation, such as EPR-steering correlations.  
\par
In their seminal paper, Einstein-Podolsky and Rosen (EPR) discussed the appearance of a ``paradox", displayed by the correlations present in some bipartite quantum states.  More recently, M. Reid and collaborators \cite{reid88,reid89,reid10} derived a number of EPR criteria that identify situations in which quantum correlations are strong enough for the EPR argument to be valid.  It has been shown that these EPR correlations are in fact equivalent to Schr\"odinger's ``steering" of quantum states \cite{schrodinger35}. 
 EPR-steering correlations were formalized recently by Wiseman et al. \cite{wiseman07} and others \cite{jones07,cavalcanti09}.  It has also been shown that steering is related to security in certain quantum cryptography protocols \cite{branciard12}. Non-steerable bipartite states can be described by  \emph{local hidden state} (LHS) models of the form \cite{wiseman07,jones07,cavalcanti09}:
\begin{equation}
P(a,b) = \sum_\lambda p_\lambda P_\alpha(a| \lambda) P_{\beta Q}(b|\lambda).
\label{eq:lhs}
\end{equation}
Here $a$ ($b$) is the outcome of measurement $\alpha$ ($\beta$) and $\lambda$ is a (``hidden") variable that labels the local state, the distribution of which is described by classical probability distribution $p_\lambda$.  $P_{\alpha}(a|\lambda)$ and $P_{\beta Q}(b|\lambda)$ are the probability distributions for outcomes $a$ and $b$ for each $\lambda$, where the subscript ``$Q$" indicates that the conditional probability  $P_{\beta Q}(b| \lambda)$ results from measurements on a quantum state.   
\subsection{EPR-Steering Criteria with variances}
Following the derivation in Refs. \cite{cavalcanti07,cavalcanti09}, we consider the sum of variances inferred about variables of system 1 via measurements on system 2:
\begin{equation}
\Delta_{\mathrm{inf}}(a)^2 + \Delta_{\mathrm{inf}}(a^\prime)^2
\label{eq:varsum}
\end{equation}
where $a$ ($a^\prime$) are the outcomes of measurement $\alpha$ ($\alpha^\prime$),  the inferred variance is
\begin{align}
\Delta_{\mathrm{inf}}(a)^2 &=  \int db P(b) \int da P(a|b) \Delta(a|b)^2 \nonumber \\
 & = \int da db P(a,b) [a - a_{\mathrm{est}}]^2,
 \label{eq:infvar}
\end{align}
$ \Delta(a|b)^2
$ is the variance of the conditional distribution $P(a|b)$, and $a_{\mathrm{est}}$ is the estimate of $a$ given outcome $b$ on system 1.    Let us assume that the correlations are described by the LHS model \eqref{eq:lhs}.  Then the inferred variance is
\begin{align}
\Delta_{\mathrm{inf}}(a)^2 & = \sum p_\lambda \int da P(a|\lambda) [a - a_{\mathrm{est}}]^2 \nonumber \\
& \equiv  \sum p_\lambda \Delta(a|\lambda)^2,
\end{align}
where we used $\int P(b|\lambda) db=1$.  Returning to expression \eqref{eq:varsum}, we have
   \begin{equation}
\Delta_{\mathrm{inf}}(a)^2 + \Delta_{\mathrm{inf}}(a^\prime)^2  =  \sum p_\lambda [ \Delta(a|\lambda)^2 + \Delta(a^\prime |\lambda)^2]. 
\label{eq:varsum2}
\end{equation}
Each $\lambda$ in the LHS model \eqref{eq:lhs} labels a quantum state.  Thus, for each $\lambda$, an uncertainty principle must be satisfied.  
Considering variables $n_{j}$ and $s_j$ ($j=1,2$), and  
using the uncertainty relation \eqref{eq:uncrel}, one arrives at the EPR-criteria 
 \begin{equation}
 \Delta_{\mathrm{inf}}(n_1)^2 + \ell^2 \Delta_{\mathrm{inf}}(s_1)^2   \geq C.
\label{eq:steercrit}
\end{equation}
Violation of criterion \eqref{eq:steercrit} thus indicates that the state is ``EPR-steerable", and cannot be represented by the LHS model \eqref{eq:lhs}. The inferred variances can be determined from  the conditional variance using Eq. \eqref{eq:infvar}.

\subsection{EPR-Steering Criteria with Shannon entropy}

Using the LHS model \eqref{eq:lhs}, it is possible to show that this probability distribution leads to \cite{walborn11a}
 \begin{equation}H(a|b) \geq \sum_\lambda p_\lambda H_{Q}(a|\lambda),
 \end{equation}
 where the conditional Shannon entropy is
 $H(a|b) =-\sum_{j} \mathcal{P}(b_j)H(a|b=b_j)$.
 Let us consider a second set of conjugate observables $\alpha^\prime$ and $\beta^\prime$, for which we arrive at an equivalent expression.  Adding the two together results in
  \begin{equation}
  H(a|b) + H(a^\prime|b^\prime) \geq \sum_\lambda p_\lambda \left \{ H_{Q}(a|\lambda) + H_{Q}(a^\prime|\lambda) \right \}.
 \end{equation}
   Each $\lambda$ in the LHS model \eqref{eq:lhs} represents a different physical realization of a quantum state.  Thus, for each $\lambda$, the state prepared should satisfy a quantum-mechanical uncertainty relation, such as the entropic criterion \eqref{eq:entuncrel}. 
 Using the normalization of $p_\lambda$, this leads directly to an entropic steering criteria:
 \begin{equation}
H(n_1|n_2) + h(s_1|s_2) \geq -\ln \ell.
\label{eq:entsteer}
\end{equation}
  Any non-steerable state can be described by the LHS model \eqref{eq:lhs}, leading directly to inequality \eqref{eq:entsteer}.  Violation of inequality \eqref{eq:entsteer} thus shows that the quantum state shared by Alice and Bob is steerable.    

\subsection{Experimental Test of EPR-Steering Criteria}
We next tested the EPR-steering inequalities. Violation of the EPR-Steering criteria \eqref{eq:steercrit} and \eqref{eq:entsteer}, calculated from the  distributions in FIG. \ref{fig:results}, are shown in FIG. \ref{fig:results2} c) and d), respectively.     In both figures, the shaded red bars show the theoretical prediction for the entangled state \eqref{eq:slitstate}. Again, the variance criteria \eqref{eq:steercrit} does not reliably identify EPR-steering for $D>2$, due principally to noise in the near-field measurements.  On the other hand, the entropic criteria \eqref{eq:entsteer} detects EPR-steering for all values of $D$.  Again, there is a considerable discrepancy between experiment and theory, particularly for $D=3,4$.  This is due to the noise in the near-field measurements, and also to the fact that the probability distribution for $N_-$ is not constant, as it is for the ideal state \eqref{eq:slitstate}.  Nonetheless, the entropic criteria  \eqref{eq:entsteer} successfully identifies EPR-steering correlations.  We thus expect the entropic entanglement criteria \eqref{eq:entcrit} and EPR-steering criteria \eqref{eq:entsteer} to be quite useful for detection of correlations in modular variables.    
\par
\section{Conclusions}
We have shown that entanglement in modular components of position and momentum variables can be detected in the spatial degrees of freedom of photon pairs.  The photons are engineered to both pass through the same slit of a D-slit aperture, and exhibit non-local interference effects.   By introducing entanglement and EPR-steering criteria based on the Shannon entropy, we are able to detect quantum correlations for $D=2,3,4$ slits.   Measurement of the modular variables is the same as the continuous $x$ or $p$ variables, and thus presents no additional experimental challenge.   {The witnesses derived here are valid for any initial bipartite quantum state.  In practice, the scaling parameter $\ell$ can be optimized to provide the best violation, or in the case that the slit separation is unknown.}  We note that the quantum correlation present in both discrete and continuous variables is similar to correlations observed in both angle (continuous) and angular momentum (discrete) \cite{leach10}.        
\par
Modular variables are a natural way to discretize continuous variables, and may prove extremely useful for continuous variable quantum information processing.  One example application is in quantum key distribution. Consider a situation analogous to the experiment presented here.  Alice and Bob could measure their respective $x$ and $p$ variables.  When they both measure $x$, they should see correlations in their $D$-dimensional discrete variables $n_1$ and $n_2$, which can be used to construct a shared key, giving $\log_2 D$ secret bits per sifted event.  Security can be checked by using a portion of the $x$ results together with the $p$ results to test for entanglement or EPR-steering \cite{branciard12}.  We expect to find similar utility in other quantum information tasks.   
\appendix
\section{Entropic Uncertainty Relation for Modular Variables}
\label{sec:uncposmom}
Consider a wavefunction of the form
\begin{equation}
\gamma(\theta) = \frac{1}{\sqrt{2 \pi}}\sum_{n=-\infty}^{\infty} c_n \exp(i n \theta). 
\label{eq:gamfun}
\end{equation}  
The sum in \eqref{eq:gamfun} is a Fourier series of the function $\gamma(\theta)$, with discrete coefficients $c_n$.   This is analogous to angle ($\theta$) and angular momentum ($n$) in quantum mechanics.  The probability distribution of continuous variable $\theta$ is $|\gamma(\theta)|^2$ and discrete variable $n$ is described by a discrete set of probabilities $|c_n|^2$, where $\sum_n |c_n|^2=1$. Refs. \cite{bialynicki75,bialynicki06} derive entropic uncertainty relations for these variables, namely
\begin{equation}
H(n) + h(\theta) \geq \ln (2 \pi).  
\label{eq:unc1}
\end{equation}
Here $H(z)$ is the discrete Shannon entropy 
\begin{equation}
H(z) = -\sum_z p_{z} \ln p_{z},
\end{equation}
where $p_z$ is the probability of  $z$,
$h(\theta)$ is the differential Shannon entropy of continuous variable $\theta$ \cite{cover}:
\begin{equation}
h(\theta) = -\int\limits_{-\pi}^{\pi} P(\theta) \ln P(\theta) d\theta,
\end{equation}
where $P(\theta)$ is the probability distribution of $\theta$.  
Note that all logarithms are to base $e$.
 This uncertainty relation is saturated for a state with one $c_n=1$ and the rest zero.  
 \par
This derivation was performed for a particular wave function.  Indeed, a probability distribution describing a mixed state can be expanded in the form 
\begin{equation}
\rho(\theta) = \sum_j a_j |\gamma_j(\theta)|^2, 
\end{equation}
where $\rho(\theta) = \bra{\theta} \hat{\varrho} \ket{\rho}$, $\gamma_j(\theta)=\langle{\theta}\ket{\gamma_j}$, $a_j\geq 0$ and $\sum_j  a_j = 1$.  Since this is a convex sum, and each $\gamma_j$ can be expanded in the form \eqref{eq:gfun}, it follows \cite{bialynicki75,bialynicki06} that the mixed state $\rho$ also obeys the entropic uncertainty relation \eqref{eq:unc1}. 
\par
Let us show that an uncertainty relation analogous to \eqref{eq:unc1} applies to the modular variables defined in Eq.  \eqref{eq:modvar}.  
Following Ref. \cite{gneiting11}, consider a wave function of the form
\begin{equation}
\Psi(x) = \sum_{n=-\infty}^{\infty} c_n \psi_n(x), 
\end{equation}  
where $\psi_n(x)$ is a normalized wavefunction centered at $n d$ and $\sum_n |c_n|^2=1$. This wave function is a superposition of wavepackets $\psi$, and might describe a particle passing through an $N$-slit aperture, for example (of course in this case all but $N$ of the $c_n$ are zero). The Fourier transform of this function gives the  wavefunction in wavevector space
\begin{equation}
\Phi(p) = \sum_{n=-\infty}^{\infty} c_n \phi_n(p), 
\end{equation}  
where
\begin{equation}
\phi_n(p) = \int\limits_{-\infty}^{\infty} e^{i 2 \pi x p} \psi_n(x) dx, 
\end{equation}  
is the Fourier transform of $\psi_n(x)$.   We are assuming that $\Psi(x)$ is a periodic comb of $\psi$-functions spaced a distance $d$ apart, so can write $\psi_n(x) = \psi_0(x-nd)$, so that $\phi_n(p)=\exp(2i \pi n d p)\phi_0(p)$.  Then
  \begin{equation}
\Phi(p) = \phi_0(p)\sum_{n=-\infty}^{\infty} c_n \exp(2i \pi n d p). 
\end{equation}  
Changing to the modular variables defined in Eqs. \eqref{eq:modvar}, we have
  \begin{equation}
\Phi(p) = \phi_0\left(\frac{m}{\ell}+s\right)\sum_{n=-\infty}^{\infty} c_n \exp(2i \pi n d [m/\ell+s]). 
\end{equation}  
It is natural to choose $\ell=d$, so that $\exp(2i \pi n \ell [m/\ell+s]) = \exp(2i \pi n \ell s)$.  If $\phi_{0}(p)\sim \mathrm{constant}$ over the range $-1/2\ell$ to $1/2\ell$, then $\phi_0(m/\ell + s) \approx \phi_0(m/\ell)$.    We have
  \begin{equation}
\Phi(p) \approx \phi_0\left(\frac{m}{\ell}\right)\sum_{n=-\infty}^{\infty} c_n \exp(2i \pi n \ell s). 
\end{equation}  
In other words, $\Phi(p) \approx \phi_0\left(m/\ell\right)g(s)$, where we define
  \begin{equation}
g(s) = \sqrt{\ell} \sum_{n=-\infty}^{\infty} c_n \exp(2i \pi n \ell s). 
\label{eq:gfun}
\end{equation}  
Let us define a variable $\theta = 2 \pi \ell s$, so that $g(\theta/2\pi\ell)$ is analogous to Eq. \eqref{eq:gamfun}.  We note that from $-1/2\ell < s < 1/2\ell$ we have $-\pi < \theta < \pi$.  Using the uncertainty relation \eqref{eq:unc1} then leads directly to
\begin{equation}
H(n) + h(s) \geq \ln \frac{1}{\ell}.
\label{eq:entuncnew}
\end{equation}
\par
By a similar argument as that outlined above for relation \eqref{eq:unc1}, uncertainty relation \eqref{eq:entuncnew} applies to all quantum states $\rho$, even those that cannot be described explicitly by wavefunction \eqref{eq:gfun}. 
The uncertainty relation \eqref{eq:entuncnew} is the main result of this section.  We will use this relation to derive tests for entanglement and EPR-steering correlations.  
 \subsection{Choice of scaling factor}
\label{sec:lmod}
\begin{figure}
  \begin{center}
 \includegraphics[width=7cm]{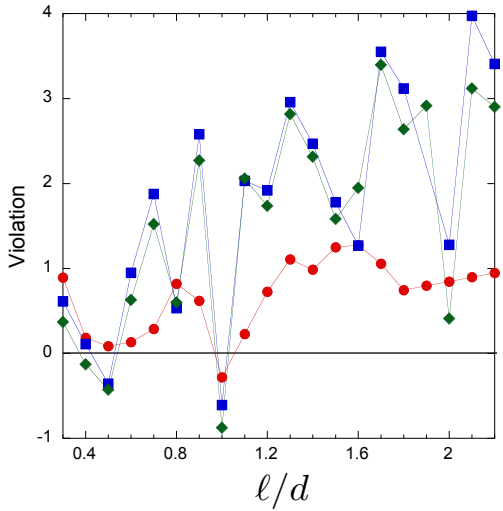}
  \caption{   
Plot of the violation of entropic criteria \eqref{eq:uncrelNs+CG} by state \eqref{eq:slitstate} as a function of $\ell/d$ for $D=2$ (red circles), 3 (blue squares), 4 (green diamonds). The strongest violations occur for $\ell/d=1$.}
\label{fig:Lmod}
\end{center}
  \end{figure}

Numerical investigation of states of the form \eqref{eq:slitstate} indicates that the optimal choice of scaling factor is $\ell=d$, the slit separation.  One can also consider $\ell \neq d$.  However, in this case it is necessary to consider a modified uncertainty relation.  This is due to the fact that for $\ell > d$ one has coarse graining in the integer part of the $x$ variable, and for $\ell < d$ one has coarse graining in the integer part of the $p$ variable.  This is equivalent to a scenario in which angular momentum is measured with precision $\delta_n > \hbar$.  Bialynicki-Birula has considered these types of coarse grained uncertainty relations in Ref. \cite{bialynicki06}.  In this case, the uncertainty relation \eqref{eq:unc1} becomes    
  \begin{equation}
H(n) + H(\theta) \geq -\ln \frac{\delta_n \delta_\theta}{2 \pi \hbar},
\label{eq:unc1CG} 
\end{equation}
where $\delta_n$ is in units of $\hbar$.  Note that both entropy functions are now discrete.  Construction of an entanglement witness using relation \eqref{eq:unc1CG} follows the same arguments outlined above.  Explicitly, we have 
   \begin{equation}
H(N_\pm) + H(S_\mp) \geq \ln \frac{\sqrt{2}}{\delta_{N_{\pm}}\delta_{S_{\mp}}\ell},
\label{eq:uncrelNs+CG} 
\end{equation}    
where $\delta_{N_{\pm}}=\ell/d$ is the precision in the $N_{\pm}$ measurements and $\delta_{S_{\mp}}$ the precision in $S_{\mp}$ measurements.   When $\ell=d$ and $\delta_{S_{\mp}}=0$, inequality \eqref{eq:uncrelNs+CG} reduces to \eqref{eq:uncrelNs+}.   Fig. \ref{fig:Lmod} shows a plot of the violation of inequality \eqref{eq:uncrelNs+CG}  by state \eqref{eq:slitstate} as a function of $\ell/d$.  All results were obtained numerically. The largest violations occur for $\ell=d$. 
\par
%%%%%%%%%%%%%%%%%%%%%%%

%%%%%%%%%%%%%%%%%%%%%%%

\begin{acknowledgements}
The authors thank L. P. Berruezo for his help in the experiment, and G. B. Lemos for useful comments on the manuscript.   Financial support was provided by Brazilian agencies CNPq, CAPES, FAPEMIG, FAPERJ, and the Instituto Nacional de Ci\^encia e Tecnologia - Informa\c{c}\~ao Qu\^antica.  
\end{acknowledgements}
%\bibliographystyle{apsrev}
%\bibliography{/Users/spwalborn/Articles/Master_Bibtex}

\end{document}